# Realization of white light cavity for ulra-short optical signal storage and Processing

**Ahmet Soysal[1], Suad sedki[2] and Alexander Pavlov[3]**

*[1] Istanbul Technical University, Turkey*

*[2] Zagazig University, Egypt*

*[3] Moscow State Engineering Physics Institute, Moscow, Russia*

*\*Corresponding author: Alexander.Pavlov1970@gmail.com*

**Abstract:** In our earlier paper we demonstrated that chirped Bragg gratings (LCBGs) can compensate for a chromatic dispersion by reflecting different wavelengths at different location along the axis of the gratings. In this paper, we make use of such a dispersion-compensating property to create a white light cavity (WLC) effect that cancels the round trip phase shift due to propagation as the optical frequency is tuned. A pair of LCBGs is used to make the cavity. To fulfill WLC condition, the dispersion profile of the LCBGs is tailored by adjusting parameters such as chirping rate, modulation depth and the length of the grating region. Numerical simulation shows WLC bandwidth is ~10GHz. We also present a method to switch the WLC effect and propose data buffer systems with a pair of LCBG−WLCs.

## References and links

1. R. H. Rinkleff, and A. Wicht, "The concept of white light cavities using atomic phase coherence," Phys. Scr. T. 118,85–88 (2005).

2. A. Rocco, A. Wicht, R.-H. Rinkleff, and K. Danzmann, "Anomalous dispersion of transparent atomic two- and three-level ensembles," Phys. Rev. A, 66, 053804 (2002).

3. A. Wicht, K. Danzmann, M. Fleischhauer, M. Scully, G. Miller, and R. H. Rinkleff, "White-light cavities, atomic phase coherence, and gravitational wave detectors," Opt. Commun. 134(1-6), 431–439 (1997).

4. M. Scalora, N. Mattiucci, G. D'Aguanno, M. C. Larciprete, and M. J. Bloemer, "Nonlinear pulse propagation in one-dimensional metal-dielectric multilayer stacks: Ultrawide bandwidth optical limiting," Phys. Rev. E 73, 016603 (2006).

5. X. Liu, J. W. Haus and S.M. Shahriar, "Modulation instability for a relaxational Kerr medium," Opt. Comm. 281, 2907-2912 (2008).

6. P. Tran, "All-optical switching with a nonlinear chiral photonic bandgap structure," J. Opt. Soc. Am. B 16, 70-73 (1999)..

7. Xue Liu, Joseph W. Haus and M. S. Shahriar, "Optical limiting in a periodic materials with relaxational nonlinearity," Opt. Exp. 17, 2696-2706 (2009).

8. Yum, H., Liu, X., Jang, Y. J., Kim, M. E., & Shahriar, S. M. "Pulse delay via tunable white light cavities using fiber-optic resonators," Journal of Lightwave Technology, 29, 2698-2705 (2011).

9. R. Fleischhaker, and J. Evers, "Four wave mixing enhanced white-light cavity," Phys. Rev. A 78(5), 051802 (2007)

10. B. J. Meers, "Recycling in laser-interferometric gravitational-wave detectors," Phys. Rev. D 38, 2317 (1988).

11. G. Heinzel, K.A. Strain, J. Mizuno, K. D. Skeldon, B. Willke, W. Winkler, R. Schilling,1 A. Rüdiger,1 and K. Danzmann, "Experimental demonstration of a suspended dual recycling interferometer for gravitational wave detection" Phys. Rev. Lett. 81 No. 25 5493 (1998).

12. Xue Liu, Shih C. Tseng, Renu Tripathi, Alexander Heifetz, Subramanian Krishnamurthy, M.S. Shahriar, "White light interferometric detection of unpolarized light for complete Stokesmetric optical coherence tomography," Opt. Comm., 284, 3497–3503 (2011).

13. S. Wise, G. Mueller, D. Reitze, D. B. Tanner, and B. F.Whiting, "Linewidth-broadband Fabry-Perot cavities within future gravitational wave detectors," Classical Quantum Gravity 21, S1031 (2004).

14. S. Wise, V. Quetschke, A.J. Deshpande, G. Mueller, D.H. Reitze, D.B. Tanner and B.F. Whiting, "Phase effects in the diffraction of light: beyond the grating equation," Phys. Rev. Lett. 95, 013901 (2005).

15. O.V. Belai, E.V. Podivilov, D.A. Shapiro, "Group delay in Bragg grating with linear chirp," Opt. Comm., 266, 512-520 (2006).

16. P. Yeh, 'Optical Waves in Layered Media', Wiley (1991).J. S. Shirk, R. G. S. Pong, F. J. Bartoli, and A. W. Snow, "Optical limiter using a lead phthalocyanine," Appl. Phys. Lett. **63**, 1880-1882 (1993).

## Introduction

White light cavity (WLC) has a broader linewidth without loss of a build-up factor than an ordinary cavity with the same finesse. Such an enhanced linewidth is an essential property to apply WLC to optical detection, sensing and communication: For example WLC increases the sensitivity enough to detect the extremely weak side band signal produced by gravity waves without restricting the detection bandwidth. For a data buffer system, the high transmission of WLC over a broad enough spectral range to encompass the data pulse spectrum has been proposes as a data buffer system. It shows an enhanced delay time bandwidth product (DBP) overcoming constraints encountered by conventional buffer systems. WLC effect in a laser cavity was proposed for hypersensitive rotation sensing wherein the sensitivity of the lasing frequency to displacement was enhanced on the order of $\sim 10^5$ higher than a conventional laser.

In WLC, the frequency dependent phase shift due to propagation delay is cancelled by tailoring a dispersion profile of the intracavity medium such that WLC condition is achieved: $n_g = 1 - L/\ell$ where $L$ is the cavity length, and $n_g$ and $\ell$ are the group index and the length of intracavity medium, respectively. In previous implementations, $n_g$ is controlled by coupling a weak probe to a strong pump in non-linear media. However, such a probe-pump interaction scheme is not applicable when we need to use a high power probe, for example, $\sim 40$ Watt probe beam is used in the Advanced LIGO interferometer and then the pump would have to be even stronger to induce dispersion. Current material technology cannot provide a non-linear medium which holds such high power beams. A passive approach to the enhancement of the bandwidth of LIGO like interferometer was attempted by using two gratings placed in parallel; however, if one consider the geometrical optical path arising from the wavelength-dependent diffraction angle as well as the additional phase change associated with the spatial phase modulation of the gratings, it is then impossible to make the variation of the phase with respect to frequency become zero. The essence of this constraint originates from the constant grating period. In ref. 错误！未定义书签。, if the grating period is a function of frequency i.e. $g \equiv g(\omega)$ rather than the constant, then the phase variation would become $d\Phi/d\omega = \beta L(\omega)/c - D tan \left[ g/\omega \right]$ implying that $d\Phi/d\omega$ could be zero with the appropriate choice of $dg/d\omega$.

Fibre interferometer configurations such as the Michelson and Fabry-Perot (FP) have been formed using chirped Fibre Bragg Gratings (FBG) acting as partial reflectors. As well as increasing the dynamic range of the interferometer, chirped FBGs are dispersive elements which can allow tuning of the response of the interferometers to measurements such as strain and temperature. In a chirped FBG, the resonance condition of the FBG varies along the FBG's length. Each wavelength is reflected from different portion of the FBG, which imparts a different group delay to the different components of the incident light. The implication of the wavelength dependence resonance position is that there is a large movement of the resonance position when the incident wavelength is changed. A chirped FBG FP can be configured in which the large dieplacement of the reflection positions in the respective FBGs forming the cavity changes in such a way that the sensitivity of the cavity can be enhanced or reduced. The FP filter response can be tailored through the extent of chirp.

In this paper, we consider a pair of linearly chirped Bragg gratings(LCBGs) in waveguide or fiber in order to create WLC effect. Thus, the grating period is not constant any more such that $d\Phi/d\omega$ can be zero. We search parameters, for example chirping rate, the length of the grating region and modulation depth, to produce the dispersion which makes $d\Phi/d\omega = 0$.

## 2. Qualitative Description

First, it is necessary to understand qualitatively how WLC condition is fulfilled by LCBG. To this end, it is instructive to review Bragg reflection from LCBG. A Bragg grating consists of index-modulated layers with periodicity. Light is partially reflected at each interface between high and low refractive index regions. When the grating period is multiple integer number of wavelength, each reflected wave is in phase resulting in high total reflection close to 100%. Consider an effective optical path length

$\Lambda_i$ ($i=1\sim3$) in the grating region $\ell$ which an input wave travels before it is fully reflected. Fig 1.(a) indicates that if the grating period changes (ex: linearly chirped gratings), $\Lambda_i$ varies with wavelength. This is an important property which allows us to produce WLC. Fig.1(b) illustrates the WLC effect in a cavity made with two LCBGs where $L$ is the cavity length. Here, we consider the case that the LCBGs have the same parameters. They are placed as displayed in Fig.1(b) such that the traveling waves inside the cavity always see the positive chirping rate. The chirping rate is tailored to make $\Lambda_i$ fulfill a resonance condition: $2\Lambda_i+L\equiv L_{eff}=m\lambda$ where $m$ is the positive integer, $\lambda$ is the wavelength and $L_{eff}$ is the effective cavity length. In doing so, we can in principle make the cavity resonate all the time though the wavelength changes.

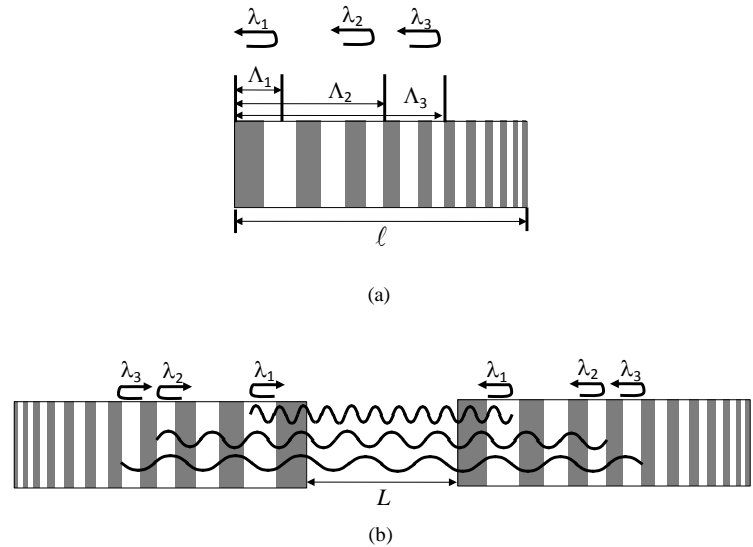

(a)

(b)

Fig.1. (a) Three different wavelengths ($\lambda_1<\lambda_2<\lambda_3$) are reflected at different locations inside the grating region (b) Schematic illustration of a typical Fabry-Perot (FP) cavity of length L formed by a pair of LCBGs.

3. Theoretical analysis

*System 1*     $L$     *System 2*

$$t_1 = |t_1| exp(j\phi_{t_1})$$

$$r_1 = |r_1| exp(j\phi_{r_1})$$

$$t_2 = |t_2| exp(j\phi_{t_2})$$

$$r_2 = |r_2| exp(j\phi_{r_2})$$

Fig.2. Schematic illustration of modeling LCBGs-WLC: the system 1&2 represent each LCBG and are separated by L to form a FP cavity.

To quantitatively analyze LCBGs-WLC, we model LCBGs with a field transmission coefficient $t_i = |t_i| exp(j\phi_{t_i})$ and a reflection coefficient $r_i = |r_i| exp(j\phi_{r_i})$ (*i*=1,2). We assume $R_i + T_i = 1$ where $R_i = |r_i|^2$ and $T_i = |t_i|^2$. Note $r_i$ and $t_i$ are complex numbers with the phase changes $\phi_{r_i}$ and $\phi_{t_i}$ resulting from the reflection and the transmission respectively. Here, $r_i$ and $t_i$ are assumed not to be constant but depend on frequency as well as physical parameters of the system. Next, as displayed in Fig.2, a pair of LCBGs (noted as system 1&2) is used to make a cavity with length *L*. The input field $E_{in}$ of the cavity would simply be related to the output $E_{out}$: $E_{out}/E_{in} = |t_1||t_2| exp\left[j\left(\phi_{t_1} + \phi_{t_2}\right)\right] / \left(1 - |r_1||r_2| exp\left[j\left(\phi_{r_1} + \phi_{r_2}\right)\right]\right) exp(2jn_0\omega L/c)$, where $n_0$ is the mean index in the cavity and $\omega$ is the angular frequency of the field. Hence, the cavity transmission $T_c \equiv |E_{out}/E_{in}|^2$ evidently becomes:

$$T_c = \frac{|t_1|^2 |t_2|^2}{1 + |r_1|^2 |r_2|^2 - |r_1||r_2| cos(\Phi_{total})} \quad (1)$$

where $\Phi_{total} = \phi_{r_1} + \phi_{r_2} + 2n_0\omega L/c$. Here we consider $T_c$ for the case in Fig.1(b). As displayed in Fig. 1(b), the traveling waves in the cavity will see the same positive chirping rate. Thus they experience the equal phase shift $\phi_r \equiv \phi_{r_1} = \phi_{r_2}$ and reflectivity $|r| \equiv |r_1| = |r_2|$. In this situation, we also found that these conditions are valid: $|t| \equiv |t_1| = |t_2|$, $\phi_t \equiv \phi_{t_1} = \phi_{t_2}$. The proof of this claim will be discussed later in detail. Therefore, $T_c$ is rewritten as: $T^2 / \left[1 + R^2 - R cos(\Phi_{total})\right]$ where $|t|^2 = T$, $|r|^2 = R$. $\Phi_{total}$ includes the term $\phi_r$ which is a function of $\omega$ and a variable system parameter $\xi$, i.e. $\phi_r(\omega, \xi)$. Note that $\xi$ is independent of frequency. The cavity is assumed to resonate at the frequency $\omega_0$ so that $\Phi_{total}(\omega_0) = 2m\pi$ (*m*: positive integer). When the frequency changes to $\omega_0 + \Delta\omega$, the phase shift from

$\Phi_{total}(\omega_0)$ is given by:

$$\Phi_{total}(\omega_0 + \Delta\omega) - \Phi_{total}(\omega_0) = 2\left(\phi_r(\omega_0 + \Delta\omega) - \phi_r(\omega_0) + \frac{n_0 \Delta\omega L}{c}\right)$$

$$\simeq \frac{2L\Delta\omega}{c}\left(n_0 + \frac{c\phi_{r1}}{L}\right) \quad (2)$$

where $\phi_{r1} = \partial\phi_r / \partial\omega\big|_{\omega=\omega_0}$. For this approximation, we consider Taylor expansion of $\phi_r(\omega,\xi)$ around $\omega_0$ keeping only the linear term $\phi_{r1}$. This linear approximation holds since we assume $\Delta\omega \ll \omega_0$. Eq.(2) indicates that if we make $\phi_{r1} = -n_0 L/c$ by adjusting $\xi$, $\Phi_{total}(\omega_0 + \Delta\omega)$ would then remain $2m\pi$. That implies that the cavity resonates for all the frequencies over a wide range of $\Delta\omega$, so-called the WLC condition.

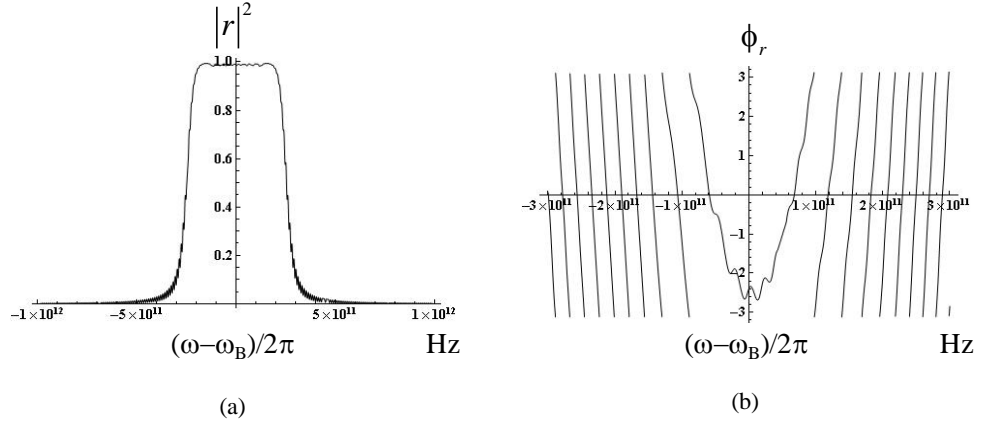

Fig.3 Numerical simulation results for (a) $|r|^2$ and (b) $\phi_r$. Note that $\phi_{r1} = -n_0 L/c$ at $(\omega - \omega_B)/2\pi = -1 \times 10^{-11}$.

In what follows, we make use of LCBG to realize a system with variable $\phi_r$. LCBG is a phase modulated index-grating. If the phase modulation occurs along z-axis, the index change is then written as $\delta n(z) \beta \cos[\theta z2\ \ (\ )]$ where $\beta$ is the modulation depth and $\theta(z) = \alpha z^2/2 + \kappa$ in terms of $\kappa$ the modulation frequency at $z=0$, $\alpha$ the chirping parameter. The analytical solution to $r$ is well-known:

$$r = \frac{\exp(j\alpha z_0^2/2)}{jk_0\beta} \frac{F\left(j\eta; 1/2; -j\alpha(\ell/2+z_0)^2/2\right) + \beta^2 k_0^2\rho(-\ell/2-z_0)F\left(1/2+j\eta; 3/2; -j\alpha(\ell/2+z_0)^2/2\right)}{\rho F\left(-j\eta; 1/2; j\alpha(\ell/2+z_0)^2/2\right) + (-\ell/2-z_0)F\left(1/2-j\eta; 3/2; j\alpha(\ell/2+z_0)^2/2\right)}$$

,

$$\rho = -\frac{F\left(j\eta; 1/2; -j\alpha(\ell/2-z_0)^2/2\right)}{\beta^2 k_0^2\rho(\ell/2-z_0)F\left(1/2+j\eta; 3/2; -j\alpha(\ell/2-z_0)^2/2\right)}$$

,

(3)

$$F(a;b;x) = \sum_{n=0}^{\infty} \frac{a_n}{b_n} \frac{x^n}{n!}; \ a_0 = 1; \ b_0 = 1; \ a_n = a(a-1)\cdots(a-n+1); \ b_n = b(b-1)\cdots(b-n+1),$$

where $z_0 = (2k-\kappa)/\alpha$ and $\eta = \beta^2 k_0^2/(2\alpha)$ in terms of $k_0 \equiv \kappa/2$. The center of the grating is assumed to be at z=0 so that LCBG is at $-\ell/2 \le z \le \ell/2$. Note that $z_0$ implies the point where a wave of the wave number $k$ propagating along z-axis is reflected and hence there is no reflection at $z > z_0$. Thus $z_0$ is relevant to the effective optical length: $\Lambda \approx z_0 + \ell/2$. Here we found $\Lambda$ depends on the frequency, which implies that each frequency experiences a different path length and hence different phase shift $\phi_r$ after the reflection from LCBG. Since $\phi_r$ depends on the design of the LCBG, it is possible to make $\phi_{r1} = -n_0 L/c$. Fig.3 illustrates $|r|^2$ and $\phi_r$ for a specific set of parameters. For illustration, we have chosen $\alpha = 277.01 \times 10^4 \, \text{m}^{-1}$, $\beta = \beta_0/2.8$, $n_0 = 1.45$ and $\ell = 0.5 \times 10^{-2} \, \text{m}$, where $\beta_0 = 0.67 \times 10^{-3}$. These parameters were chosen to fulfill the WLC condition $\phi_{r1} = -n_0 L/c$ for $L \equiv L_0 = 0.640876 \times 10^{-2} \, \text{m}$ at $\omega_0 = \omega_r \equiv \omega_B - 2\pi \times 10^{11}$ where $\omega_B = 2\pi \times 1.9355 \times 10^{14}$ corresponding to $\lambda = 1550 \, \text{nm}$. To see WLC effect, we will detune frequency around $\omega_r$ and $\omega_r$ is chosen to make sure $|r|^2 \approx 1$ for the particular detuning range. Since for $\omega < \omega_B$ $\partial \phi_r / \partial \omega < 0$ shown in Fig.3(b), WLC effect would be created within $\omega < \omega_B$. Also, we simulate $|t|^2$ and $\phi_t$ for $\alpha = \pm 277.01 \times 10^4 \, \text{m}^{-1}$ (everything else same) using the analytical solution to $t$ presented in ref **错误！未定义书签。**. From the simulation results (not shown in figure), we found $|t|^2$ and $\phi_t$ are the same for different $\alpha$ such that $|t| \equiv |t_1| = |t_2|$ and $\phi_t \equiv \phi_{t_1} = \phi_{t_2}$ in the cavity displayed in Fig.1(b).

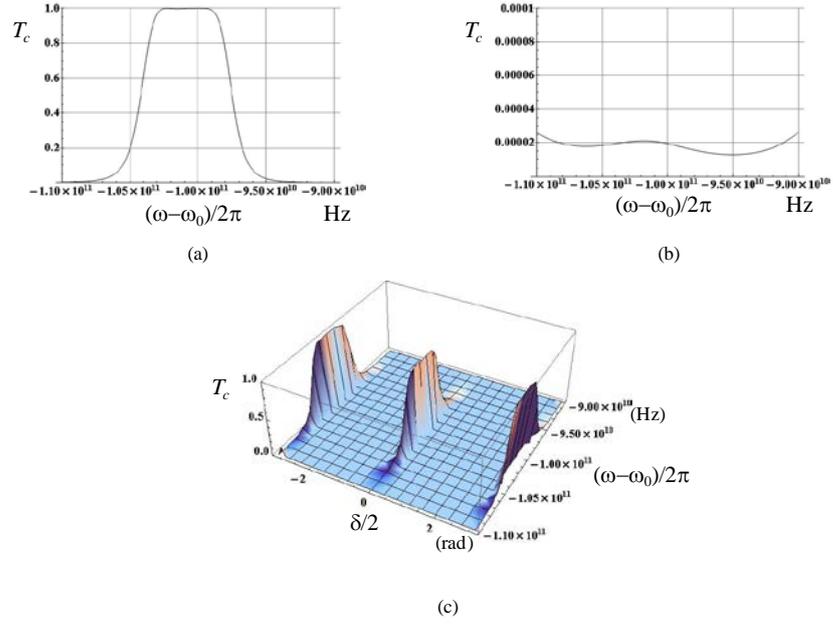

(a)

(b)

(c)

Fig.4. Numerical simulations for $T_c$ with (a)$\delta$=0, (b) $\delta$=$\pi$, and (c) $-\pi/2<\delta<\pi/2$. The parameters for LCBG are the same as used in Fig.3.

Fig.4 displays the transmission of the cavity with the length $L_0$. For illustration, we insert $|r|^2$ and $\phi_r$ obtained from Fig.3 to Eq.(1). In Fig.4(a), it is evident that the cavity shows WLC effect with the full width at half maximum (FWHM) ~2.9GHz. Note that $T_c$ is asymmetrical around $\omega_r$ because $\phi_r$ does not change linearly. For the applications in optical communication considered later, in particular data buffer systems 错误！未定义书签。, we discuss how to eliminate the WLC effect. To this end, it is instructive to plot $T_c$ with the additional phase $\delta$ so that $\Phi_{total}=2\phi_r+2n_0\omega L_0/c+\delta$. We will discuss methods to induce $\delta$ in detail later. Fig.4(b) indicates $T_c\simeq 0$ for $\delta=\pi$ meaning that WLC effect is off. Fig.4(c) illustrates $T_c$ for $-\pi<\delta/2<\pi$. In Fig 4(c), it is evident that WLC effect appears when $\delta=2q\pi$ ($q$ is integer).

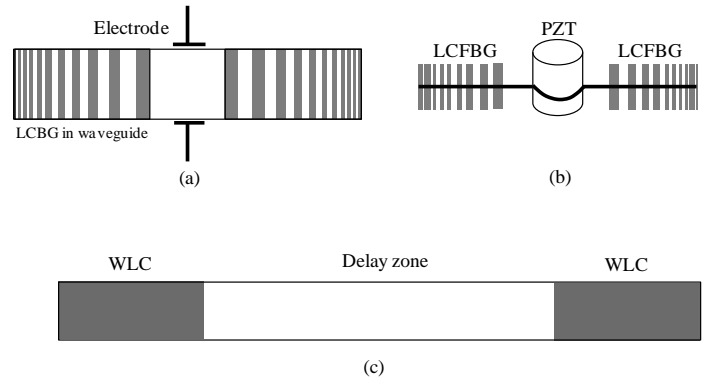

Fig 5. Schematic illustrations for the switchable WLC: (a) voltage through the electrodes produces electro-optic effect resulting in the phase Π. As such, light undergoes the additional phase shift δ=2Π, (b) PZT induces $\Delta L$ giving the additional phase $\delta = 2n_0\omega\Delta L/c$ to the propagating wave. (c) Data buffer system with the two WLCs placed in series.

It is worth to discuss the practical method to induce $\delta$ in LCBG WLC. Fig 5(a) and (b) illustrate two different systems to realize a switchable WLC: LCBG WLC in fiber where a piezoelectric actuator (PZT) changes the cavity length to $L_0+\Delta L$ so that $\delta = 2n_0\omega\Delta L/c$, and LCBG WLC in waveguide structure where the electro-optic effect due to the voltage through the electrodes gives Π so that $\delta = 2\Pi$. It is important to note that $\Delta L$ and Π are independent of wavelength but are determined by the voltage on the PZT and through the electrodes, respectively. Hence, light will experience the same phase shift $\delta$ regardless of the wavelength and also $\delta$ is adjustable by changing the voltage. As presented in Fig. 5(c), this switchable WLC allows us to make a data buffer system in a manner analogous to the previously proposed buffer systems. When a pulse enters from left, there is no voltage on the WLC resulting in $\delta = 0$ i.e. WLC effect is on. As soon as the pulse transmits through the WLC, the voltage is activated to induce the additional phase. With the particular value $\delta \equiv \delta_0$ suitable for the elimination of the WLC effect, the pulse will see the high reflectivity of LCBG/ Bragg reflector (BR) and thus remain trapped between them. When the voltage is off, we can then release the trapped pulse.